\documentclass[prx,superscriptaddress,preprint]{revtex4-1}
\usepackage[utf8]{inputenc}
\usepackage{amsmath,amsfonts,amssymb}
\usepackage{graphicx}
\usepackage{subfig}
\usepackage{multirow}
\newcommand{\etal}{et al }

\newcommand{\eg}{e.g. }
\newcommand{\ie}{i.e. }
\newcommand{\spt}{space-time}
\newcommand{\reffig}[1]{Fig. \ref{#1}}

\newcommand{\kphotonics}{Photonics Laboratory, King Abdullah University of Science and Technology \emph{(KAUST)}, Thuwal 21534, Saudi Arabia}
\newcommand{\kfupm}{King Fahd University of Petroleum and Minerals \emph{(KFUPM)}, Dhahran 31261, Saudi Arabia}

\begin{document}
\title{A possible approach on optical analogues of gravitational attractors}
\author{\firstname{Dami\'an P.} \surname{San-Rom\'an-Alerigi}}
\affiliation{\kphotonics}
\email[email: ]{damian.sanroman@kaust.edu.sa}
\author{\firstname{Ahmed} \surname{Benslimane}}
\affiliation{\kphotonics}
\author{\firstname{Tien K.} \surname{Ng}}
\affiliation{\kphotonics}
\author{\firstname{Mohammad} \surname{Alsunaidi}}
\affiliation{\kphotonics}
\affiliation{\kfupm}
\author{\firstname{Boon S.} \surname{Ooi}}
\affiliation{\kphotonics}

\begin{abstract}\emph{
In this paper we report on the feasibility of light confinement in orbital geodesics on stationary, planar, and centro-symmetric refractive index mappings. Constrained to fabrication and [meta]material limitations,  the refractive index, $n$, has been bounded to the range:  $0.8\leq n\left(\vec r \right)\leq 3.5$. Mappings are obtained through the inverse problem to the light geodesics equations, considering trappings by generalized orbit conditions defined \emph{a priori}. Our simulation results show that the above mentioned refractive index distributions  trap light in an open orbit manifold, both perennial and temporal, in regards to initial conditions. Moreover, due to their characteristics, these mappings could be advantageous to optical computing and telecommunications, for example, providing an on-demand time delay or optical memories. Furthermore, beyond their practical applications to photonics, these mappings set forth an attractive realm to construct a panoply of celestial mechanics analogies and experiments in the laboratory.}\newline
Published at Optics Express {\bf 21}, 8298--8310, 2013. DOI: 10.1364/OE.21.008298.
\end{abstract}
\maketitle

\section{Introduction\label{sec.intro}}

\begin{figure}[htb]
\centering
	\begin{tabular}{c c}
		\subfloat[~]{\includegraphics[width=.4\textwidth]{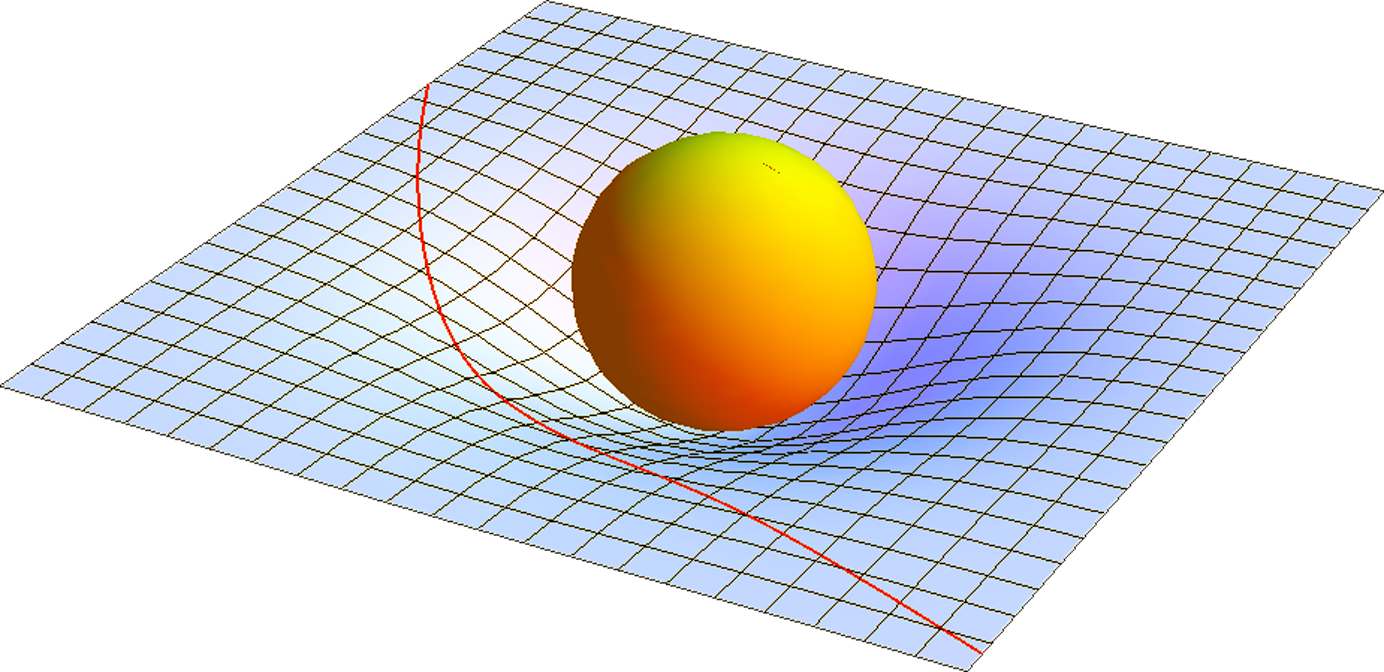}\label{fig.bending}} &
		\subfloat[~]{\includegraphics[width=.4\textwidth]{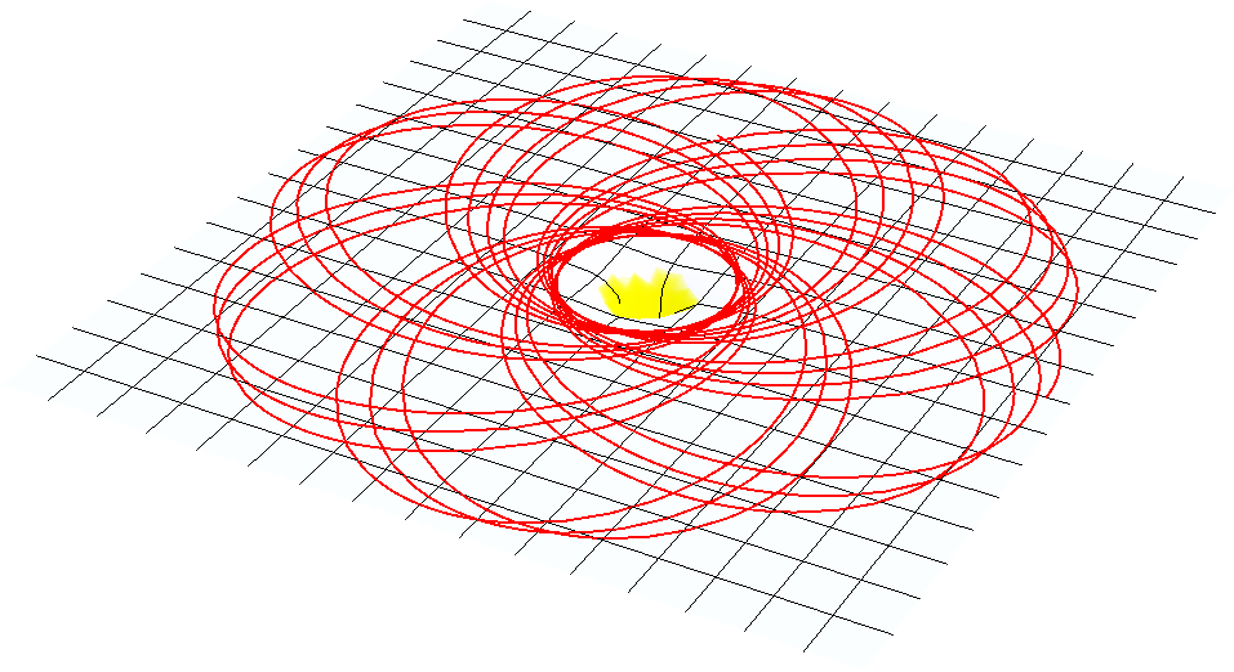}\label{fig.bhole1}} \\
		\subfloat[~]{\includegraphics[width=.4\textwidth]{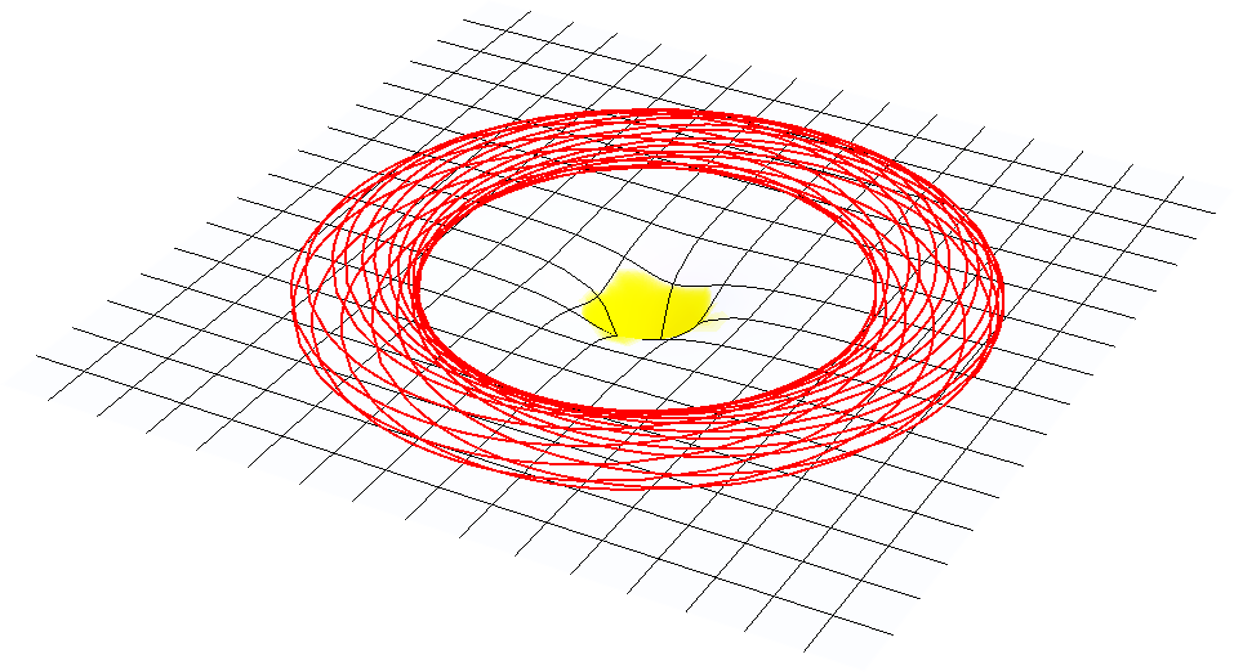}\label{fig.bhole2}} &
		\subfloat[~]{\includegraphics[width=.4\textwidth]{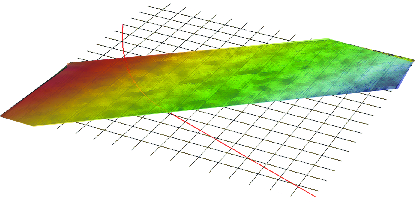}\label{fig.indbending}}
	\end{tabular}
	\caption[Comparing sidereal and laboratory frame geodesics.]{Comparison of sidereal and laboratory frame geodesics for light beams in (a)  the vicinity of a star,  (b) and (c) spinning black hole at different initial conditions, and (d)  the laboratory across an heterogeneous monotonic increasing refractive index medium.\label{fig.bholes}}
\end{figure}

Einstein's general relativity theory allows us to understand the dynamics of mass and massless particles as they travel through \spt. From it, we learn that matter  and  geometry are inextricably related; the latter transfigures the metric of \spt, and  this alteration in turn modifies the geodesics, the  paths of  particles, resulting in a manifold of relativistic effects, \emph{v. gr.} the apparent shift in the position of the stars due to the \emph{gravitational lensing effect}, depicted in \reffig{fig.bending}, where photons traveling near a massive star are deflected from their otherwise straight path due to the curvature of \spt; or trapping in multifarious orbits around black holes with respect to the initial conditions and \emph{shape} of the gravitational attractor \cite{Levin2008,Hloe2010,Muller2011}, see Figs. \ref{fig.bhole1} and \ref{fig.bhole2}.

Keeping in mind the previous illustrations, let us momentarily examine  a more mundane experiment. Imagine a photon moving in a region of space filled with some monotonically increasing refractive index. What we observe as the photon moves in the medium is that its path describes a curve like the one depicted in \reffig{fig.indbending}. If the gradient of the refractive index is tailored properly, it would be possible to build an optical analogy to gravitational bending (see \reffig{fig.bending}) for example. This leads us to wonder whether it is possible to fabricate an optical analogue to other sidereal geodesics by means of a differential control on the index of refraction.

The latter analogy reveals a connection between both examples, a weft where both realms [sidereal and laboratory] can be understood as manifestations of the same mathematical ground, albeit significantly different physics, \emph{differential geometry}. Einstein's general theory of relativity tells us that geometry and gravity are  linked by means of the metric tensor of \spt; in optics the intricacy of the relation between geometry and the physical properties of materials (permittivity and permeability) has been evinced by miscellaneous methods, most based on differential geometry  \cite{Leonhardt2008a,Chen2010,Pendry2006,Schurig2006}.

These methods can be used to show that the optical example of differential light bending  can be understood in terms of geometry, essentially by bending the space on which the light propagates, hence the metric is  related to the electromagnetic tensors \cite{Leonhardt2008a, Pendry2006}. This realization paved the way for  the theoretical realization of optical cloaks, super-lenses, optical illusion devices and recently  black-hole analogues in moving refractive index perturbations \cite{Pendry2006,Schurig2006,Leonhardt1999,Leonhardt2000,Brevik2001,Liu2009,Ergin2010,Lai2009,Kwon2009}.

Concerning trapping and light confinement in particular, different methods have been proposed\cite{Hloe2010,Genov2009,Piwnicki2001,Cacciatori2010}. Of particular interest is Genov \etal study, which presented an isotropic inhomogeneous refractive index mapping which was able to confine light mimicking a closed orbit around a black hole \cite{Genov2009}.  As the authors noted, such mapping could be implemented on a dielectric photonic crystal, composed of air-gaps in Gallium-Arsenide.

In our work, we posit theoretical traps that provide an optical analogue to celestial mechanics to realize light confinement, trapping  in  a region of space, by means of two-dimensional stationary refractive index mappings, which could be implemented under current technological and photo-refractive [meta]materials, constraint to optical frequencies, which could be easily integrable on all-optical-chips \cite{Sanroman2012,Zakery2003,Kang2009,Sanroman2013}. The mathematical and physical background to make  these effects  possible could bring forth an exciting ground to test sidereal mechanics in the laboratory; and provide the key to enable  a miscellany planar optical systems that are of great interest for diverse photonic applications, ranging from time delays, to temporal optical memories, or random resonators, to name a few;  thence its significance.

\section{Deriving the model\label{sec.model}}

We regard  a model equation based in the paraxial approximation of non-linear electrodynamics, by looking into a medium where the index of refraction can be modulated in space and time by some smooth varying function $f$, \eg in photo-refractive media index of refraction is a real valued function $f$ of the incident light intensity $I$; therefore $n(\vec x)=f(I(\vec x))$, where $\vec x$ is the \spt~quadrivector which corresponds to the coordinates $\{x^0: x^1,x^2,x^3\}$. In this representation $x^0$ corresponds to the time coordinate, and the rest to the usual three-dimensional space. In this context the photon's path is governed by the geometry of the space the light is propagating on, which in general has the metric tensor   $g_{ij}$,  and the infinitesimal  length element $ds^2=g_{ij}dx^i dx^j$.  Therefore, if we parameterized the path followed by light with reference to the proper time $\tau$, \ie $\vec x(\tau)$, we can analytically derive the path equations by means of Lagrange-Euler equations, whit a Lagrangian of the form: 
\begin{equation} \label{eq.lagrange}
	L=\frac{1}{2}\left[g_{ij}\frac{\partial x^{i}}{\partial \tau}\frac{\partial x^{j}}{\partial \tau}\right],
\end{equation}
\noindent where the derivatives are taken in respect to the affine parameter $\tau$, \ie the \emph{proper time}.

The physical description behind Eq. \eqref{eq.lagrange} is relevant to optics, as we set forth earlier in the introduction. If  the metric were indeed related to the electromagnetic tensors, permittivity and permeability, then it could be possible to write the light-path in the paraxial approximation as a function of either the refractive index $n=f(x^0,\vec r)=\sqrt{\varepsilon(\vec r,t)\mu(\vec r, t)}$, or the metric $ds$. Hence, it could be possible to use Fermat's principle, where Eq. \eqref{eq.lagrange} would take the  role of an \emph{optical Lagrangian}, \ie the solution to the variational problem which states that given a refractive index $n(x^0,\vec r)$ the path light follows is given by the extrema of:
\begin{equation} \label{eq.maximan}
	\delta \tau=\delta \int\limits_{x_1}^{x_2}ds=0,
\end{equation}
\noindent where the integration is done between the arbitrary points $x_1^j$ to $x_2^j$, with $j=0,\ldots, 3$, and $\delta$ indicates the functional maximization. Recursively, the solution to Eq. \eqref{eq.maximan} is the Lagrange-Euler Eq. \eqref{eq.lagrange}. Meanwhile, observe that in the integral of Eq. \eqref{eq.maximan} the term $ds$ takes into account the metric by virtue of the covariant metric tensor $g_{ij}$ of the curved space:
\begin{equation} \label{eq.nds}
	ds=\sqrt{g_{ij}dx^i dx^j}.
\end{equation}

Yet, we are still to determine the relation between the metric tensor $g_{ij}$ and the refractive index $n$. This can be achieved by means of conformal mappings, where it is possible to build the relation between the metric and the electromagnetic properties of a [meta]material \cite{Leonhardt2008a,Chen2010,Pendry2006}. Specifically, if the line element of \spt~  takes the form $ds^2 = g_{00}dx^0dx^0 - g_{ii}dx^i dx^i$,  it is possible to map the curved \spt~ to an isotropic, non-dispersive and non-absorbing medium as \cite{Leonhardt2008a,Genov2009}:
\begin{equation}\label{eq.ncurvedmetric}
	n=\varepsilon=\mu=\sqrt{\frac{{-g}. g^{ij}}{g_{00}}},
\end{equation} 
\noindent where $g=\det{(g_{ij})}$. Furthermore if we consider Minkowski's metric of special relativity, the tensor $g$ has the entries:
\begin{equation}
	g=diag(g_{00},g_{11},g_{22},g_{33})=diag(1,-1,-1,-1),
\end{equation} 
\noindent then the  refractive index $n=n(x^0,\vec r )$ can be found using Eq. \eqref{eq.ncurvedmetric}. And consequently, the length element $ds$ of Eq. \eqref{eq.maximan} can be written as:
\begin{equation}
	ds^2=\frac{1}{n^2}dx^0dx^0-dx^idx^i, \quad i=1,2,3
\end{equation}

\subsection{Path equations and the inverse problem}

Accordingly, from  the Lagrange-Euler \eqref{eq.lagrange} we can derive the  light-path, which in cylindrical coordinates takes the form:
\begin{align}
	\ddot r&= r \dot\phi^2+\frac{\dot t^2 \partial_r n}{n^3}\label{eq.npath1},\\
	\ddot \phi&= \frac{-2 r \dot r \dot\phi+\frac{\dot t^2 \partial_{\phi}n}{n^3}}{r^2}\label{eq.npath2},\\
	\ddot \phi&= \frac{\dot t^2 \partial_z n}{n^3}\label{eq.npath3},\\
	\ddot t&= \frac{\dot t \left(\dot t \partial_t n+2 \left(\dot z \partial_z n+\dot\phi  \partial_{\phi}n+\dot r \partial_r n\right)\right)}{n}\label{eq.npath4},
\end{align}
\noindent where all derivatives marked by $\dot{\square}$ and $\ddot{\square}$ are with respect to the affine parameter $\tau$, \eg $\dot r=\partial r/\partial \tau$. Equation \eqref{eq.npath1} through Eq. \eqref{eq.npath3} describe the parametrized path of a light beam traveling in a space-time with such geometry or equivalent refractive index.

We will describe light being trapped, when the geodesics are perpetually bounded to some region of space, for as long as the material properties remain unchanged. In this sense a light trapping geodesic is one where any of the following conditions arise:
\begin{enumerate}
	\item $\dot r, \dot\phi  \rightarrow 0$,\label{pre.1} 
	\item $\dot r \rightarrow 0$ and $\phi_{min}\leq\phi\leq\phi_{max}$,\label{pre.2}
	\item $\dot \phi \rightarrow 0$ and $r_{min}\leq r \leq r_{max}$,\label{pre.3}
	\item $r_{min} \leq r \leq r_{max}$ and $\dot \phi_{min}\leq\dot \phi\leq\dot \phi_{max}$,\label{pre.4}
\end{enumerate}
\noindent for all $\tau \geq \tau^*$; where $\tau^*$ marks the moment at which the initial conditions for trapping are met. The first condition is only possible if the refractive index $n$ goes to  infinity, or if the refractive index co-moves with the wave \cite{Piwnicki2001,Cacciatori2010}. The other three cases can be realized in \emph{stationary} refractive index \cite{Genov2009,Ni2011}. Thus, they can be imposed as general conditions to solve the inverse problem to Eq. \eqref{eq.npath1} through Eq. \eqref{eq.npath4}.

The inverse problem begins by defining a set of \emph{a priori} conditions that describes a generalized trapping scenario. If we impose premises \ref{pre.1} to \ref{pre.4}, described earlier, the path equations reduce to new set of equations where the affine parameter is time itself; whence we can rework Eq.   \eqref{eq.npath1} through Eq. \eqref{eq.npath4} into polar coordinates:
\begin{align}
	\ddot r&= \frac{n r \dot \phi^2-2 \dot r \dot \phi \partial_\phi n-\dot r^2 \partial_r n+r^2 \dot \phi^2 \partial_r n}{n},\label{eq.npath5}\\
	\ddot \phi&= -\frac{2 n r \dot r \dot \phi-\dot r^2 \partial_\phi n+r^2 \dot \phi^2 \partial_\phi n+2 r^2 \dot r \dot \phi \partial_r n}{n r^2}.\label{eq.npath6}
\end{align}

Since we are interested in the  general stationary case, we impose the fourth condition above, ensuing  $0 \leq \dot r \leq \dot r_{max}$ and $\dot \phi_{min}\leq\dot \phi\leq\dot \phi_{max}$, on  Eq. \eqref{eq.npath5} and Eq. \eqref{eq.npath6}, and  find a non-singular refractive index distribution, $n(r, \phi)$,  that satisfies them. Notice that we require the refractive index to be  smoothly varying and $C^2$ continuous, \ie $\nabla n \cdot \vec e^i \leq M$, where $M$ is a real valued number, and $\vec e^i$ a vector in a give direction in \spt. The general solution to this problem, however, is complex and a unique solution is not guaranteed \cite{Sanroman2012,14,31}.

The latter being said, the problem can be simplified further if we recall that a trapping orbit occurs when the radius $r$ is bounded to a region of space, \ie $r_{min} \leq r \leq r_{max}$, for all $\phi$; whence $\dot r(r_{max})=\dot t(r_{min})=0$, and $r$ can be written in terms of the angle $\phi$ or the time $t$. Hence a natural solution to the inverse problem is an autonomous refractive index that varies with the radial variable, \ie $n=n(r)$. This refractive index map resembles the centro-symmetric characteristics of gravitational fields, where, as pointed out independently by Genov and Ni \cite{Genov2009,Ni2011}, light would behave similarly to that traveling in curved space-time. Such distribution can be constrained to the conditions listed previously.
  
This result is significant in terms of fabrication where novel photo-refractive [metal]materials could be used. The refractive index in these composites can be modulated by the intensity of the change-inducing beam; which can be controlled by shaping its transversal profile \cite{Sanroman2012,Zakery2003,Kang2009,Sanroman2013}. Accordingly several families of refractive index distributions can be achieved, \eg Gaussian, Laguerre, or Bessel, which are all centro-symmetrical.

\subsection{System stability and dynamic equillibrium}

Since the refractive index is a function of the radius,  Eq. \eqref{eq.npath5} and Eq. \eqref{eq.npath6} can be rewritten in the form \eqref{eq.refseg}:
\begin{align}
	\ddot x&=-h(x,\dot x) - g(x),\label{eq.refseg}\\
	\ddot r&=\left(r^2\dot\phi^2-\dot r^2 \right)\partial_r\ln{n}+r\dot\phi^2,\label{eq.npath7}\\
	\ddot \phi&= -2\dot r \dot \phi\frac{n +r\partial_r n}{n r},\label{eq.npath8}
\end{align}
\noindent
the system's symmetry now allows us to integrate Eq. \eqref{eq.npath8}, hence:
\begin{equation}
	\dot \phi = \frac{\ell}{\mu r^2}, \quad \mu=n^2,
\end{equation}
\noindent
where $\ell$ is a constant of integration. This equation is reminiscent of orbital mechanics, with the sole exception that $\mu$, the mass in celestial mechanics, varies with the radius. Substituting the latter result into Eq. \eqref{eq.npath7}, and rearranging yields:
\begin{equation}
	\ddot r = \left(\frac{\ell^2}{n^4r^4}r^2-\dot r^2\right)\partial_r\ln n+\frac{\ell^2}{n^4r^4}r.\label{eq.npath9}
\end{equation}

Equation \eqref{eq.npath9} arrangement allows us to calculate the \emph{ratio of external energy supply}, which is a measure of the equilibrium stability of the system \cite{Jordan2007,Verhulst1990}; simplifying we get:
\begin{equation}
	\mathcal{E}_t=\frac{\mathrm{d} \mathcal{E}}{\mathrm{d}  t} = -\dot r h(r,\dot r) = \dot r \left(\frac{\ell^2}{n^4r^2}-\dot r^2\right)\partial_r\ln n.
\end{equation}

\begin{figure}[ht]
	\centering
	\includegraphics[width=10cm]{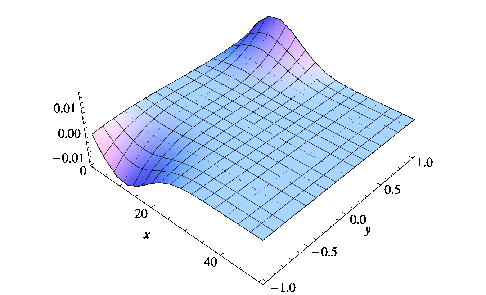}
	\caption{\emph{Ratio of external energy supply}, $\mathcal{E}_t$, for a radial symmetric system, where $x=r$ and $y=\dot r$. Notice that there exists a $x_c$, critical radius, at wich for all phase space pairs $\{x,y | x\geq x_c\}$ $\mathcal{E}_t=0$, which implies that the light geodesics do not describe an orbital motion.\label{fig.dedt}}
\end{figure}

By virtue of Bertrand's theorem \cite{Genov2009,Zarmi2002}, we know that $\dot r \neq 0$, for all time $t$; then from premises \ref{pre.3} and \ref{pre.4} it follows that $\dot r_{min} \leq \dot r \leq \dot r_{max}$, where $\dot r_{min} < 0$; and from the definition of $n(r)$ as a monotonous decreasing function it follows that $\partial_r\ln n\leq 0$. Therefore, as the radius decreases, the radial velocity goes to its minimum and $\mathcal{E}_t\leq 0$, therefore the system approaches the equilibrium point and the amplitude of the path decreases. On the other hand, as the radius increases, the velocity tends to a maximum, and  $\mathcal{E}_t\geq 0$, thus the system is driven away from the equilibrium point and the amplitude of the path increases. In Fig. \ref{fig.dedt} we show $\mathcal{E}_t$ for a Gaussian like refractive index described in section \ref{sec.single}. Observe that in the light confinement scenario we will see an oscillatory behavior in $\mathcal{E}_t$.

To conclude we study the stability of the orbits. by  means of  Lyapunov's stability theory for the governing Eq. \eqref{eq.npath9},  and derive the constraints applicable to the refractive index mapping to attain stable orbital paths \cite{Genov2009}. First, let us rewrite  Eq. \eqref{eq.npath9} as:
\begin{equation}\label{eq.npath10}
\begin{aligned}
	\dot r &= \varkappa,\\
	\dot\varkappa &=(r^2\beta^2 -\varkappa^2)\chi(r) + r\beta^2,
\end{aligned}
\end{equation}
\noindent
with $\beta=\dot\phi$ and $\chi(r)=\partial_r\ln n$. Then, if we introduce small perturbations $r=r_o + \delta$ and $\varkappa =\varkappa_o + \epsilon$,  expand $\chi(r)$ in a Taylor series around $r_o$, and drop all higher order terms for $\delta$ and $\epsilon$; we can rewrite Eq. \eqref{eq.npath10} as a linear system of equations for the perturbation terms:
\begin{equation}\label{eq.npath11}
\begin{aligned}
\left(  \begin{matrix} \dot \delta \\ \dot \epsilon \end{matrix}\right) =& \left(  \begin{matrix} 0 & 1\\ W & 2V \end{matrix}\right)\left(  \begin{matrix}  \delta \\  \epsilon \end{matrix}\right),\\
V = -\varkappa_o\chi_o, \quad &W=2\chi_o\beta^2r_o+\chi'_o(r_o^2\beta^2 - \varkappa_o^2),
\end{aligned}
\end{equation}
\noindent where $\chi_o$ and $\chi'_o$ are the linear terms of the Taylor expansion of $\chi(r)$ around $r_o$. Ensuing the eigenvalues of the matrix above are given by:
\begin{equation}
\lambda_{1,2}=V \pm \sqrt{V^2+W}.
\end{equation}

The geodesics will be stable if the real part of the \emph{eigenvalues} of Eq. \eqref{eq.npath11}, $\Re\{\lambda_{1,2}\}$, are negative or equal to zero. 
Recall that $0\leq \beta \leq \lim_{r\to\infty} n(r)^{-1}$, and $\chi(r)\leq 0$; and therefore the condition for Lyapunov stability is reached if:
\begin{equation}
	{-\frac{\partial}{\partial r}\chi(r)}\bigg|_{r=r_o}\varkappa_o \leq 0, \qquad V^2+W \leq 0,\label{eq.stability}
\end{equation}

This constrained is met for all points $r$ near the stability point $r_o=0$, provided the radial velocity meets $\dot r_o \leq 0$.  In general, since the refractive index is monotonous decrescent, the system will be Lyapunov stable at  points where $\dot r \leq 0$, see, for example, the phase-space plots in Figs. \ref{fig.4b} and \ref{fig.8b}. In addition, it is interesting to analyze whether the system will be Lyapunov stable under the conditions necessary to sustain a circular orbit. In this case $\varkappa_o = 0$, and $r_o = a$, constant, for all time $t$; consequently $V=0$, $W\leq 0$, and  the second term in the equality \eqref{eq.stability} is imaginary, provided that $\chi(r)\leq 0$; thus resulting in $\Re \{ \lambda_{1,2} \}=V=0$. These conditions, however, result in a different family of refractive indexes. In our study we focus on Gaussian-like refractive index distributions, feasible through photo-refraction, where $\varkappa \neq 0$, ergo the system will not describe closed circular orbits.

In summary, light confinement will occur as long as the conditions cited earlier apply. Under them the system will oscillate around the equilibrium point as depicted by the \emph{ratio of external energy supply}. In the next section we will present the confinement geodesics resulting from these refractive index distributions. We will show that the geodesics in this system mimic those of a spinning black hole and other celestial objects, thus reproducing the paths of light traveling in curved space.

\section{Results and discussion\label{sec.results}}

\begin{figure}[ht]
		\centering
		\begin{tabular}{c c c}
		\subfloat[$n_a=3.0, \quad n_c=0.8$]{\includegraphics[width=.3\paperwidth]{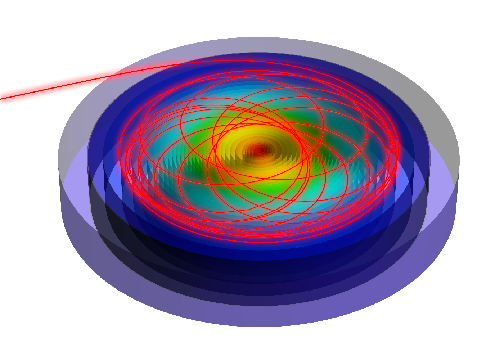}\label{fig.trapa}
		}&
		\qquad  &
		\subfloat[$n_a=3.3, \quad n_c=0 .5$]{\includegraphics[width=.3\paperwidth]{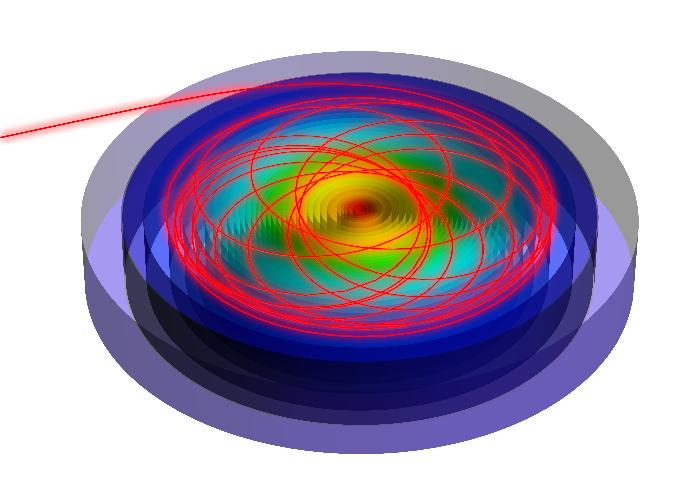}\label{fig.trapb}}
		\\
		\subfloat[$n_a=2.8, \quad n_c=1.0$]{\includegraphics[width=.28\paperwidth]{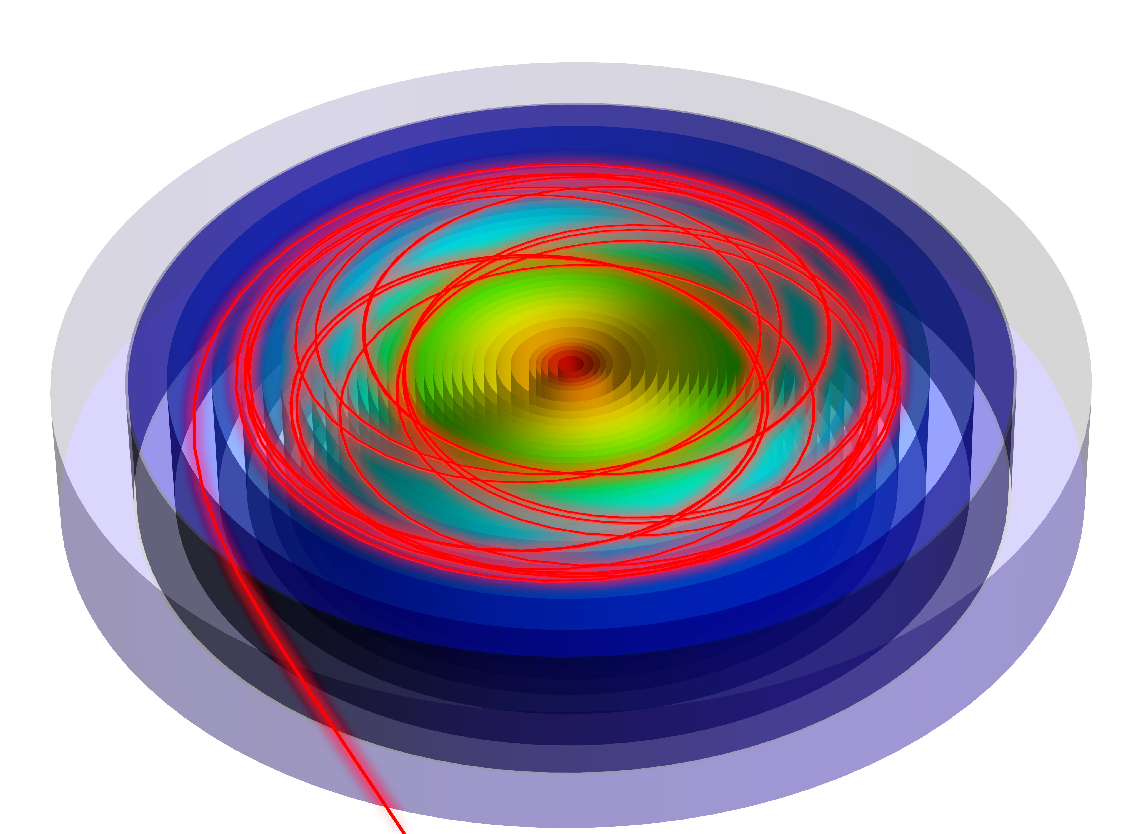}\label{fig.trapc}}
		&
		\qquad  &
		\subfloat[$n_a=2.7, \quad n_c=1.1$]{\includegraphics[width=.3\paperwidth]{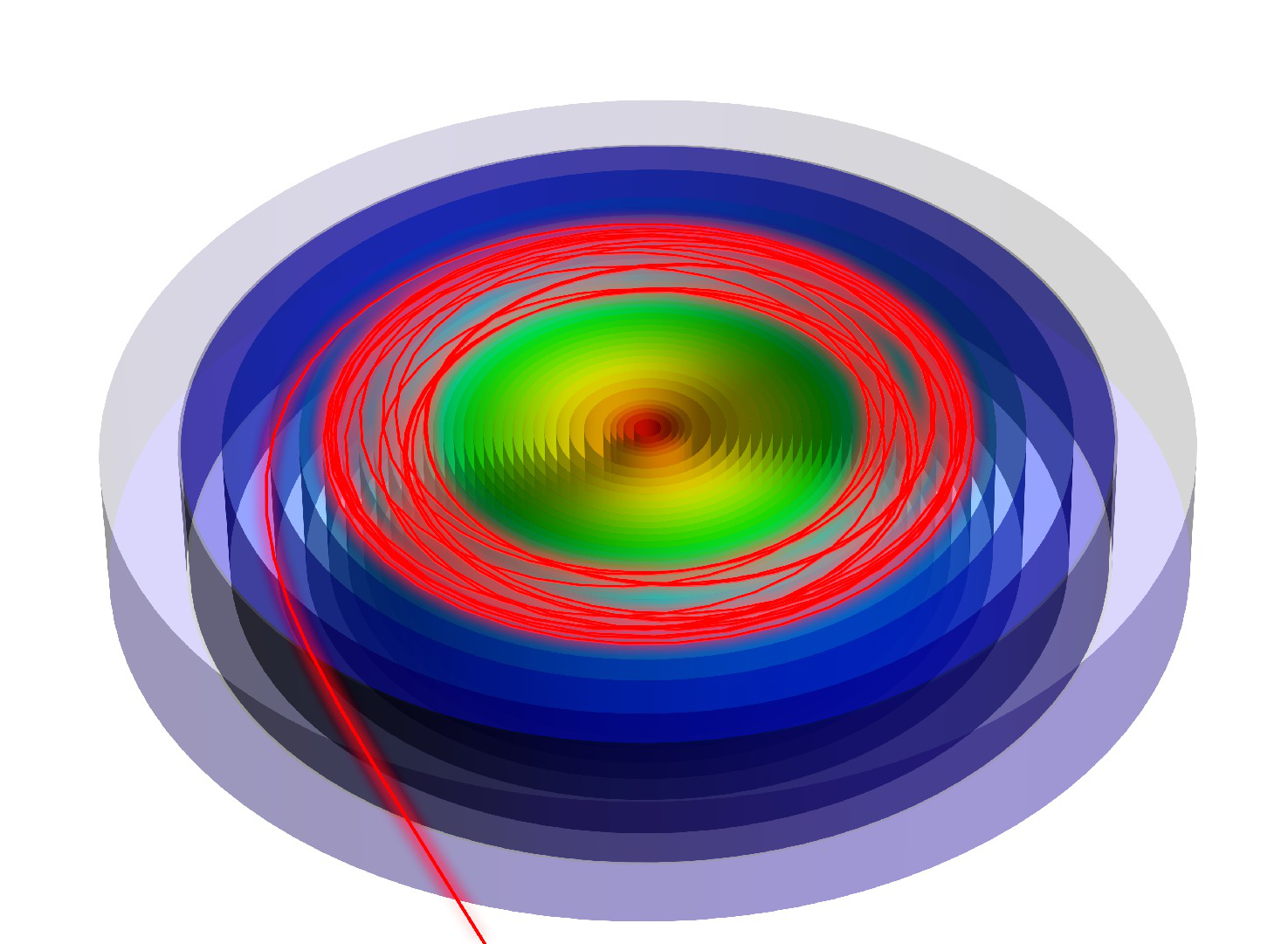}\label{fig.trapd}}
		\\
		\multicolumn{3}{c}{\subfloat{\includegraphics[width=.3\paperwidth]{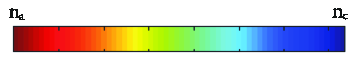}}}
		\end{tabular}
		\caption{Light path dynamics for Gaussian refractive index shape for different $n_a$ and $n_c$. The maximum value of $n$ in all cases is $3.8$ at the center of the device. $\sigma$ is constant for all results.  Notice the increment in the internal radius $r_{min}$ as $n_c$ increases, and the  orbit width reduces.}\label{fig.traporbits}
\end{figure}

In this section we explore three refractive index distributions based on  Gaussian profiles. The first two  mimic light trapped in the vicinity of a black hole, and the third case that of a particle moving in-between two planets.

\subsection{Guassian distribution, single attractor\label{sec.single}}

A Gaussian refractive index distribution has the form:
\begin{equation}\label{eq.singlea}
n=n_a e^{-r^2/\sigma^2}+n_c.
\end{equation}

Then making it possible to constrain the maximum and minimum of the refractive index distribution to feasible values, \eg $0.8 n \leq 3.8$. Hence, given a range of values for the background and maximum refractive index, $n_c$ and $n_a$, respectively,  we can evaluate different light confinement scenarios, see \reffig{fig.traporbits}.

\begin{figure}[ht]
	\centering
	\includegraphics[width=6.5cm]{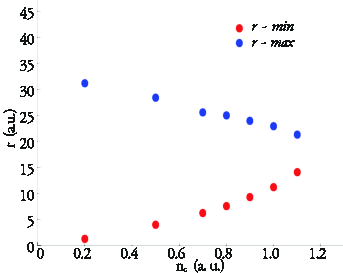}
	\caption{Inner and outer radius as a function of the background refractive index $n_c$. As it increases we observe that the outer radius diminishes, while the inner increases.}\label{fig.radii}
\end{figure}

As discussed in the previous section the orbital paths are open, while their shape and stability depend significantly on the initial conditions, $n(r)$ and $\{r_o, \dot r_o, \phi_o, \dot \phi_o\}$. As portrayed by \reffig{fig.traporbits} different initial conditions will result in  new orbit manifolds with distinct characteristics, albeit maintaining a common morphology. As it can be seen from these figures the orbital paths are highly sensitive to the modifications in the refractive index and the initial conditions; a change in $\sim 0.5 \%$ in $n_c$ alters the orbit, as seen from the plot of radii extrema in Fig. \ref{fig.radii}.

\begin{figure}[ht]
	\centering
	\begin{tabular}{c c c}
		\subfloat[]{\includegraphics[width=.4\textwidth]{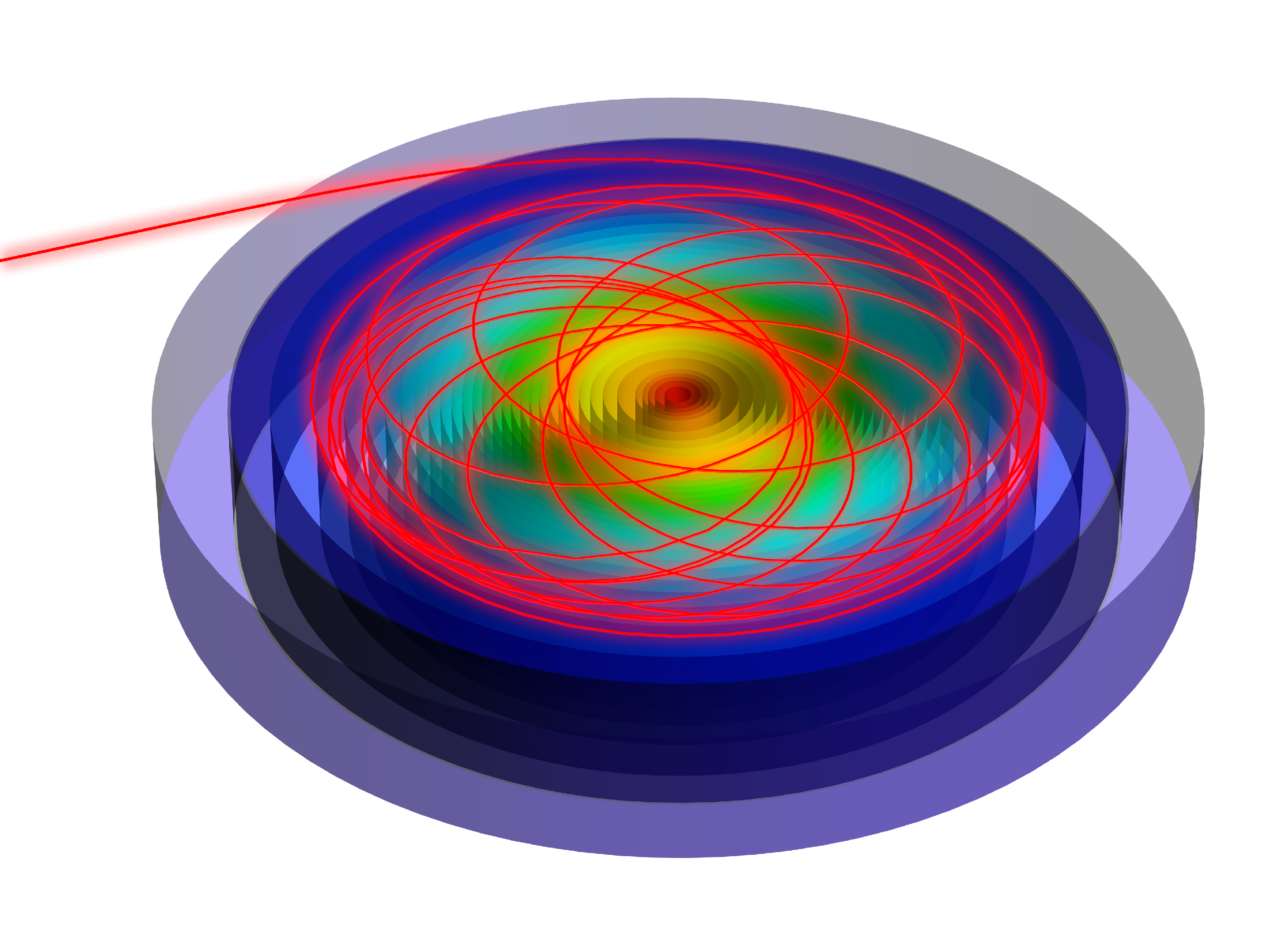}} &
		\qquad \qquad &
		\subfloat[]{\includegraphics[width=.3\textwidth]{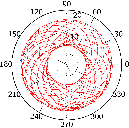}}
	\end{tabular}
	\caption{Confinement orbit for light traveling in a Gaussian attractor where $n_a=3.0$ and $n_c=0.8$. (a) Shows a schematic representation of the planar trapping device where the refractive index increases towards the center, where red indicates maximum refractive index $n=3.8$, and light blue $n=0.8$; (b) polar-2D path as calculated by the simulation routine.}\label{fig.singletrap1}
\end{figure}

\begin{figure}[ht]
	\begin{tabular}{c c c}
		\subfloat[]{\includegraphics[width=.4\textwidth]{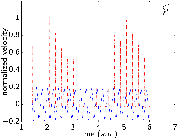}\label{fig.4a}}&
		\subfloat[]{\includegraphics[width=.3\textwidth]{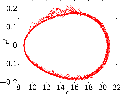}\label{fig.4b}}&
		\subfloat[]{\includegraphics[width=.3\textwidth]{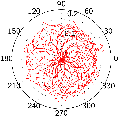}\label{fig.4c}}
	\end{tabular}
	\caption[phase space]{Phase space for Gaussian-like refractive index map with $n_a=3.0$, $n_c=0.8$. The plots show: (a) radial $\dot r$ and angular $\dot \phi$ velocities, (b) radial phase space, and (c) radial velocity as a function of the angle (c).\label{fig.gauss1phase}}
\end{figure}

In \reffig{fig.singletrap1} we show the 2D open orbit for a refractive index map where $n_c=0.8$ and $n_a=3.0$. As detailed in \reffig{fig.4a}, the radial velocity describes an oscillatory form, which indicates that the radius follows a similar motion, swinging between two extrema, as expected in a confine and open orbital motion. The latter can be observed in greater detail by analyzing the phase-space, plotted in \reffig{fig.4b}; notice that when the velocity $\dot r$ reaches zero the orbit radius $r$ reaches an extrema; for a closed single loop the amplitude of the velocity oscillation would be zero, and the phase space would become a horizontal line at $\dot r=0$. The behaviour of the angle versus the radial velocity, and consequently to the radius movement, is described by \reffig{fig.4c}, where the absolute value of the radial velocity is plotted against the angular position, which results in a periodic motion. Observe that in the limit at every angle there would be a time at which the radial velocity would be zero, which is typical of open orbits.

\subsection{Mexican hat distribution\label{sec.mexhat}}

We study a variation on the Gaussian trap to understand the potential modifications on the trapping orbits when large modifications $\sim 20\%$ of the refractive index take place. The mathematical description of the \emph{mexican hat} is:
\begin{equation}\label{eq.mexhat}
n=(n_a-\frac{n_d r^2}{\sigma^2}) e^{-r^2/\sigma^2}+n_c
\end{equation}

\begin{figure}[ht]
	\centering
	\begin{tabular}{c c c}
		\subfloat[]{\includegraphics[width=.4\textwidth]{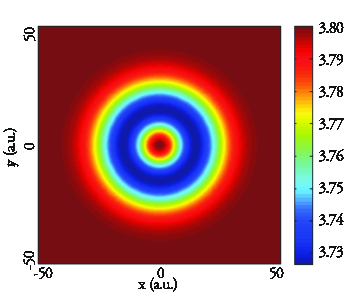}} &
		\qquad \qquad &
		\subfloat[]{\includegraphics[width=.3\textwidth]{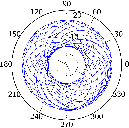}}
	\end{tabular}
\caption{(a) Schematic of Mexican hat like refractive index map where $n_a=3.0$ and $n_c=0.8$ and depression is $n_d=0.2$; (b) depicts the resulting trapped orbit in polar coordinates as calculated by the simulation routine.}\label{fig.singletrap2}
\end{figure}

\begin{figure}[ht]
	\begin{tabular}{c c c}
		\subfloat[]{\includegraphics[width=.4\textwidth]{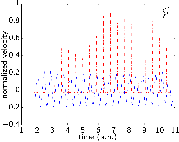}\label{fig.8a}}&
		\subfloat[]{\includegraphics[width=.3\textwidth]{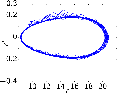}\label{fig.8b}}&
		\subfloat[]{\includegraphics[width=.3\textwidth]{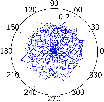}\label{fig.8c}}
	\end{tabular}
	\caption[phase space]{Phase-space for a \emph{mexican hat} refractive index map with$n_a=3.0$, $n_c=0.8$. The plots show: (a) radial $\dot r$ and angular $\dot \phi$ velocities, (b) radial phase space, and (c) radial velocity as a function of the angle (c).\label{fig.gauss2phase}}
\end{figure}

Using the same conditions as in the previous analysis we set $n_c=0.8$, $n_a=3.0$, and set $n_d=0.2$. The results, shown in Figs. \ref{fig.singletrap2} and \ref{fig.gauss2phase}, display an orbit whose appearance resembles that of the regular Gaussian trap studied earlier. However, a closer look reveals that the trap confinement area is $\sim 2\%$ slimmer, and the oscillatory frequency of the radial and angular velocities, \reffig{fig.8a},  increases by $\sim 30\%$. Whereas the radial phase-space, \reffig{fig.8b}, shows a smaller eccentricity compared to its Gaussian counterpart, and the absolute value of the radial velocity versus the angle, \reffig{fig.8c}, describes the typical behaviour of an open orbit.

These variations in the trapping conditions could potentially be used to differentiate trapping in transient optical memories to allocated binary data.

\subsection{Double attractors\label{sec.twosuns}}

Interested in observing phenomenologically trapping in a binary system we evinced a planar refractive index map resulting from the superposition of two Gaussian functions,
\begin{equation}
n=n_a(e^{-(r-r_{off1})^2/\sigma^2}+e^{-(r-r_{off2})^2/\sigma^2})+n_c.
\end{equation}

\begin{figure}[ht]
	\centering
	\begin{tabular}{c c}
		\subfloat[$n_a=2.7, \quad n_c=1.1$]{\includegraphics[width=.3\paperwidth]{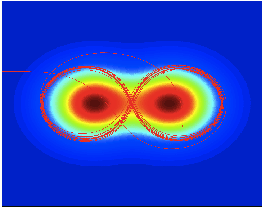}} &
		\subfloat[$n_a=2.9, \quad n_c=.9$]{\includegraphics[width=.3\paperwidth]{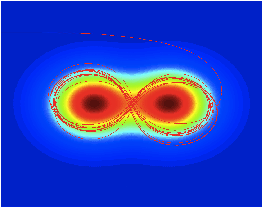}}
	\end{tabular}
	\caption{Geodesics for light in a binary system. Observe that trapping also occurs for certain initial conditions at specific background and maximum refractive index.}\label{fig.twosuns}
\end{figure}

The resulting orbits are shown in \reffig{fig.twosuns}. A binary system, however, presents a relatively higher sensitivity to modifications in the refractive index, where a distortion of $\Delta n > 0.03$ modifies the trapping orbits, in some cases releasing the light beam.Yet, it is possible to confine the light beam to an orbit which mimics the behaviour of Newton's three body problem, where, as expected,  small perturbations result in sundry possible solutions to the path equation, \ie chaos.

\subsection{The space-time equations, trapping}\label{sec.eikonal}

\begin{figure}[ht!]
	\centering
	\includegraphics[width=.8\textwidth]{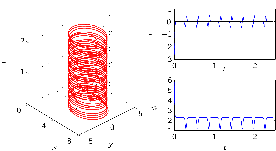}
	\caption[phase space]{Phase space of the trapping orbit for $n_a=3.0$, $n_c=0.8$. The plots show: (a) radial $\dot r$ and angular $\dot \phi$ velocities, (b) radial phase space, and (c) radial velocity as a function of the angle.\label{fig.eikonal}}
\end{figure}

Briefly, we show that the same Gaussian solution explored in section \ref{sec.single} fulfills the trapping conditions in the full eikonal space-time, Eq. \eqref{eq.npath1} through Eq. \eqref{eq.npath3}. The results are shown in \reffig{fig.eikonal}.

In this case we observe a similar behaviour to that of the Gaussian trap in the stationary approximation, the difference in the amplitude values is due to normalization. An oscillatory behaviour in the radius and angular velocities is described in Figs. 10(b) and 10(c); as it would be expected from a centro-symmetrical refractive index mapping.

\subsection{Material realization}

Recent advancement in photo-refractive materials \cite{Sanroman2012,Zakery2003,Kang2009,Sanroman2013} could enable the construction of the light confinement devices discussed. For example, a prototype Gaussian trap could be fabricated  by stacking different photo-refractive [meta]materials like $As_xS_{1-x}$ in multiple layers. Recalling that in photo-refraction the optical properties of the material, permeability and permittivity, change as a function of the light intensity, allowing the fabrication of a continuously varying refractive index as the one required by a centro-symmetrical  attractors. Moreover, since photo-refractivity is elastic, \emph{i.e.} reversible, it allows one to modify the refractive index of the layers combined in time, thence enabling or disabling the trap at will.

\section{Conclusions}

We have demonstrated that, under the paraxial description of light in heterogeneous media, it is possible to use centro-symmetrical refractive index distributions as attractors for light, which are able to perform light trapping or confinement in open orbits.  These mappings are bound within fabrication constrains, and due to the paraxial approximation they are several times larger than the mean beam width and wavelength, $\sim 50$ times larger. The proposed mappings have potential applications to transient optical memories, delays, concentrators, random cavities, and beam stirrers, to cite a few; which could enable the next generation of photonic integrated circuits, PICs, processors and computation systems.   Due to their sensitivity to the change in the refractive index these devices could be employed as a sensor platform. Furthermore, since they are drawn in close analogy to celestial phenomena they could be used as laboratory setups to test a large variety of effects in celestial mechanics.

\section*{Acknowledgments}

The authors wish to acknowledge David Ketcheson, Carlos Argaez, Pedro Guerra, and Jose Alcantara-Felix for their suggestions, corrections and valuable discussions while preparing this manuscript.

\bibliographystyle{plainnat}

\end{document}